\documentstyle[12pt,epsf]{article}
\font\twelve=cmbx10 at 15pt
\font\ten=cmbx10 at 12pt

\def\build#1_#2^#3{\mathrel{\mathop{\kern 0pt#1}\limits_{#2}^{#3}}}

\def\vect#1{\overrightarrow{#1\kern 1pt}\kern-1pt}

\newcommand{\beq}[1]{\begin{equation}\label{#1}}
\newcommand{\eeq}{\end{equation}}
\newcommand{\ba}{\begin{array}}
\newcommand{\ea}{\end{array}}
\newcommand{\beqan}{\begin{eqnarray*}}
\newcommand{\eeqan}{\end{eqnarray*}}

\def\pr{Phys.\ Rev.\ }
\def\prl{Phys.\ Rev.\ Lett.\ }
\def\pl{Phys.\ Lett.\ }
\def\np{Nucl.\ Phys.\ }

\def\VV{{\cal V}}

\newcommand{\dis}{\displaystyle}

\newcommand{\ds}{\displaystyle \strut}

\def\vpc{\vbox to 0,8cm{}}

\epsfverbosetrue

\begin{document}

\begin{titlepage}

\begin{center}

\renewcommand{\thefootnote}{\fnsymbol{footnote}}

{\ten Centre de Physique Th\'eorique\footnote{
Unit\'e Propre de Recherche 7061} - CNRS - Luminy, Case 907}

{\ten F-13288 Marseille Cedex 9 - France }

\vspace{1,2 cm}

{\twelve TOP QUARK PRODUCTION\\
IN EXTENDED BESS MODEL}

\vspace{0.3 cm}

\setcounter{footnote}{0}
\renewcommand{\thefootnote}{\arabic{footnote}}

{\bf R. CASALBUONI}\footnote{Dipartimento di Fisica, Univ. di Firenze,
I.N.F.N, Firenze}, {\bf P. CHIAPPETTA}, {\bf D. DOMINICI}$^1$, \\
{\bf A. FIANDRINO}\footnote{and Universit\'e de Provence, Marseille}
{\bf and R. GATTO}\footnote{D\'epartement de Physique Th\'eorique,
Univ. de Gen\`eve, Gen\`eve}

\vspace{1,5 cm}

{\bf Abstract}

\end{center}

 We study top production at Tevatron collider in the extended BESS
model, which is an effective lagrangian
parametrization of a dynamical symmetry
breaking of the electroweak symmetry. The existence of a colored
octet of gauge vector bosons can increase top production at
a rate still consistent with recent experimental data and
lead to distorsions in the transverse momentum spectrum of the top.

\vspace{1,5 cm}


\bigskip

\noindent Number of figures : 4

\bigskip

\noindent April 1995

\noindent CPT-95/P.3176

\noindent UGVA-DPT 1995/04-887

\bigskip

\noindent anonymous ftp or gopher: cpt.univ-mrs.fr

\end{titlepage}

               %%%%%%%%%%%%%%%%% 1 %%%%%%%%%%%%%%%%%%

\section{Introduction}

Very recently the CDF \cite{cdfn} and D0 \cite{d0n}
collaborations have established the existence of the
top by observing a signal consistent with $t \bar t$ decay
into $W^+W^-b \bar b$ with a statistical significance
of $4.8 \sigma$ and $4.6 \sigma$ respectively.
The observed topologies rely on the decay modes of the W
 bosons leading to dilepton events or to events with
isolated leptons and more than three jets. To
suppress background in the second mode b quark tagging
is mandatory. The CDF measurement of the top quark
mass is $m_t= 176 \pm 8 \pm 10$ GeV with a cross section
of $6.8^{+3.6}_{-2.3}$ pb, whereas D0 collaboration  has
measured   a mass $m_t= 199^{+19} _{-21}\pm 22$ GeV with a
production cross section of $6.4\pm 2.2$ pb.
 From  LEP, which is sensitive to
heavy particles through radiative corrections, it has been possible
to extract a top mass in the same mass range \cite{lep}($m_t= 178 \pm
11 ^{+18} _{-19}$ GeV), reinforcing our confidence that the top is
around 180 GeV.

Since the top has a mass of  the order of the electroweak symmetry
breaking scale, it is not unreasonable to expect that it may be
strongly coupled to the symmetry breaking dynamics and
therefore the possibility that top production may give an insight
on new physics beyond the SM has to be considered.

This sensitivity to new physics has been recently investigated in
the context of walking technicolor \cite{mt}. The pseudo-goldstone
color octet $\eta_T$, characterized by a small decay constant, can
significantly increase and even double the
top production rate at Tevatron collider if it lies in the mass
range
$M_{\eta_T}\sim 400-500 \hbox{GeV}$ \cite{eichten}. The existence of
a color octet technirho boson, or contact terms from an effective
lagrangian involving dimension six operators, may  significantly
modify the transverse momentum distribution of the top quark and the
distribution of the decay $W$ boson \cite{hill}. Concerning the
total cross
section for top production it will also lead to  significant
deviations from SM expectations.
Since present measurements of top
production rate do not indicate a
strong deviation from SM
expectations, they will put constraints on this type of new physics.

In what follows we will restrict on the consequences of a strong
dynamical electroweak symmetry scheme called the BESS model
\cite{bess}, which provides for a general frame avoiding specific
dynamical assumptions. More precisely we will consider the extended
BESS \cite{exbess} - based on a chiral structure $SU(8)_L\times
SU(8)_R$ -  which includes as a special case  the
low energy phenomenology of the one family technicolor model. This
model leads to the existence of pseudo-goldstone bosons and new
vector and axial vector resonances. The colored pseudo-goldstone
bosons
$\pi^{\alpha}_8$ and the massive partners of the gluons, denoted as
$V^{\alpha}_8$, might significantly modify top production. The purpose
of this work is to perform such a quantitative study.

Our extended-BESS calculations are only aimed at phenomenological
studies, in the spirit of an effective lagrangian approach, including
spin-1 bosons of the strong electroweak sector. The problem of an
underlying field-theoretic dynamics, such as technicolor and extended
technicolor models, is beyond the BESS model approach, which aims at
being more general though of a more limited role. For the
non-extended BESS, specialization to technicolor and extended
technicolor was discussed in ref. \cite{bess-tech}. Much work by
different authors has been devoted during the last years to improve
on the original technicolor ideas \cite{technicolor}, after the soon
appreciated difficulty of flavor changing neutral currents and the
related fermion mass problem \cite{mass}. We have already referred to
the proposal of walking technicolor \cite{mt} to induce bigger
techni-condensates and allow for large fermion masses with extended
technicolor at the needed large scale. The existence of an acceptable
model with realistic properties is however still an unanswered
question, complicated by the overall uncertainties related to
non-perturbative strong dynamics. For the weak isospin-conserving
technicolor the large mass difference between the top and the bottom
quark would require weak-isospin violation at the extended
technicolor level, which immediately raised up some doubts on the
possibility of satisfying the experimental limits on the parameter
$\rho$, in ordinary or in walking technicolor schemes. This lead in
1989 to consideration of schemes of strong extended technicolor
\cite{strong-tech}, to have large top masses and satisfy the limits
on $\rho$, for the values that were accepted at that time. Apart from
the theoretical uncertainties typical of non-perturbative strong
dynamics, a fine-tuning close to the line of criticality appeared
necessary in those models \cite{strong-tech}, suggesting additional
scalars of light mass \cite{light}. For the vertex $Zb\bar b$ (see
our discussion in Section 5 below) one may also expect smaller
corrections in these cases. As we have said, we shall entirely
refrain from speculating on particular underlying dynamical models
and use BESS as a free effective model, only subject to existing
experimental data. In particular this will concern the discussion of
the possible direct couplings (called $b_S$ and $b_S'$ in Section 4
below) with would be hints at extended technicolor, when interpreted
within such dynamical schemes. We must add that recent investigations
have dealt with models of non-commuting extended technicolor, where
the gauge boson of the extended technicolor generating the mass of
$t$ has non trivial weak-$SU(2)$ transformation (with implications on
$Zb\bar b$
\cite{chiv}). Finally, as a further hint to the richness of dynamical
schemes which have been recently proposed, we will also refer to a
new class of technicolor models which incorporate a new strong
dynamics mostly coupled to $t$ and $b$, and broken by technicolor
\cite{colortop}. Such schemes lead to consequences such as for
instance "top-pions" and new testable phenomenological effects, as
described in ref. \cite{colortop}.

The paper is organized as follows. We will first recall in section
2 some basics of the $SU(8)$ BESS model. We study in section 3 the
effect of a color octet of technipions on the top production at
Tevatron collider. Section 4 is devoted to the consequences of the
existence of an octet of massive colored vector bosons. We discuss
the numerical results at Tevatron energy in section 5 and we conclude
in section 6.

               %%%%%%%%%%%%%%%%% 2 %%%%%%%%%%%%%%%%%%

\section{The $SU(8)$ BESS model}

In a dynamical scheme for electroweak symmetry breaking an
initial global invariance symmetry characterized by a group $G$,
here
$G=SU(8)_L\times SU(8)_R$, is spontaneously broken into a subgroup
$H$, here $SU(8)_{L+R}$.

Assuming that the information for the fermion mass
mechanism can be embedded into effective Yukawa couplings
between ordinary fermions and pseudo-goldstones, their
mass spectrum can be derived from the one loop effective
potential which includes also the ordinary gauge interactions.
One gets for the masses of the colored states \cite{pseudo}:
\beq{1}
\ba{ll}
M^2(\pi_8^{\alpha\pm})
&=
\frac{\dis \Lambda^2}{\ds 4\pi^2v^2}\left[m_t^2+m_b^2+
\frac{9}{2}v^2g_s^2\right] \\[3mm]
M^2\left(\frac{\dis \pi_8^{\alpha}+\pi_8^{\alpha
3}}{\ds \sqrt{2}}\right)
&=
\frac{\dis \Lambda^2}{\ds 2\pi^2v^2}\left[m_t^2+
\frac{9}{4}v^2g_s^2\right] \\[3mm]
M^2\left(\frac{\dis \pi_8^{\alpha}-\pi_8^{\alpha
3}}{\ds \sqrt{2}}\right)
&=
\frac{\dis \Lambda^2}{\ds 2\pi^2v^2}\left[m_b^2+
\frac{9}{4}v^2g_s^2\right] \\[3mm]
\ea
\eeq
where $\Lambda$ is a cut-off, here  taken equal to $\Lambda= 2\
\hbox{TeV}$, $g_s$ the strong coupling constant and $v=246\
\hbox{GeV}$. For  $\Lambda= 2 \ \hbox{TeV}$, the typical mass scale
of the colored technipion is  $\sim 1 \hbox{TeV}$.

After performing the $SU(3)_C\times SU(2)_L\times U(1)_Y$
gauging, the strong part of the lagrangian reads:
\beq{2}
\ba{ll}
{\cal L}_s
=&
-\frac{\dis v^2}{\dis 4}\left\{b Tr\left(\sqrt{2}g_s G_{\mu}^{\alpha}
T_8^{\alpha} - g" V_{\mu 8}^{\alpha} T_8^{\alpha}\right)^2
\right.\\[3mm]
&\left. +c Tr\left( g" A_{\mu 8}^{\alpha} T_8^{\alpha}\right)^2+
dTr\left( g" A_{\mu 8}^{\alpha} T_8^{\alpha}\right)^2\right\}
\ea
\eeq
where $G_{\mu}^{\alpha}$ $(\alpha = 1,...,8)$ is the octet of
$SU(3)_C$ and $V_{\mu 8}^{\alpha}\  (\hbox{resp.} \ A_{\mu
8}^{\alpha})$ the new color octet of vector (resp. axial vector)
resonances. The $SU(8)$ generators satisfy the algebra :
\beq{3}
[T^A,T^B] = if^{ABC} T^C \ \hbox{with} \ Tr(T^A
T^B)=\frac{1}{2} \delta^{AB}
\eeq
The eigenstates are obtained, after diagonalisation of the
mass matrix, in the basis $(G_{\mu}^{\alpha}, V_{\mu
8}^{\alpha}, A_{\mu 8}^{\alpha})$ :
\beq{4}
\frac{v^2}{4}
\pmatrix{2b g_s^2 & -\sqrt{2}b g_s g" & 0 \cr
-\sqrt{2}b g_s g" & b g"^2 & 0 \cr
0 & 0 & (c+d) g"^2 \cr}
\eeq
leading to the physical states characterized by the masses :
\beq{5}
\ba{ll}
M_{\cal G}^2
&=
0 \\[5mm]
M_{\VV_8}^2
&=
\frac{\dis 1}{\dis 4}b g"^2 v^2
\left[1+\frac{\dis 2g_s^2}{\ds g"^2}\right]\\[5mm]

M_{{\cal A}_8}^2
&=\frac{\dis 1}{\dis 4}
(c+d) g"^2 v^2\\[5mm]
\ea
\eeq
The physical bosons ${\cal G}_{\mu}, {\cal V}_{\mu 8}$ are
related to $G_{\mu}^{\alpha}$ and $V_{\mu 8}^{\alpha}$ by :
\beq{6}
\ba{ll}
{\cal G}_{\mu}
&=
\cos\xi\cdot G_{\mu} + \sin\xi\cdot V_{\mu 8}  \\[3mm]
{\cal V}_{\mu 8}
&=
-\sin\xi\cdot G_{\mu} + \cos\xi\cdot V_{\mu 8}\\[3mm]
\ea
\eeq
with
\beq{7}
\tan\xi = \frac{\dis \sqrt 2 g_s}{\ds
g"}
\eeq
\noindent Notice that we get a massless color octet vector
boson, the physical gluon, as it should be.

\noindent  The coupling of the gluon to a quark antiquark pair
is given by :
\beq{8}
g_s\cos\xi \gamma_{\mu}\frac{\dis \lambda^a}{2}
\eeq
We must have : $g_s'=
g_s\cos\xi = g_{QCD}$.

\noindent The coupling to the physical colored massive gauge boson
is given by:
\beq{9}
-g_s\sin\xi \cdot\gamma_{\mu}\frac{\dis \lambda^a}{2}
\hskip0.5truecm  \hbox{i.e} \hskip0.5truecm  -\frac{\dis
\sqrt{2}g^{'2}_s}{\ds \sqrt{g"^2-2g_s^{'2}}}
\gamma_{\mu}\frac{\dis \lambda^a}{2} \eeq

\section{Top pair production through a color-octet pseudo
goldstone boson $P_8$}

The color octet pseudo goldstone boson is expected to
decay predominantly into the heaviest kinematically allowed
fermion pair or into $g g$ through the anomaly.

\noindent The decay into $t\bar{t}$ is calculated from the
lagrangian :  \beq{10}
{\cal L}_y = m_t \bar{t}_R e^{\dis \frac{\dis 2i}{\ds
v}(T_8^{\alpha} \pi_8^{\alpha}+ T_8^{\alpha 3}  \pi_8^{\alpha 3})}t_L
+
\hbox{h.c}
\eeq
leading to a partial width:
\beq{11}
\Gamma(P_8^0 \rightarrow t\bar{t}) =\frac{1}{2\pi} \left(
\frac{m_t}{v}\right)^2 M_{P_8^0} \sqrt{1-\frac{\dis 4m_t^2}
{\ds M_{P_8^0}^2}}
\eeq
where $P_8^0=\frac 1 {\sqrt 2} (\pi_8-\pi_8^3)$.
The amplitude for $\pi_8^{\alpha} \rightarrow g g$ is calculated
{}from the Adler-Bell-Jackiw triangle anomaly \cite{anomaly}.

\noindent One gets
\beq{12}
\Gamma(\pi_8^{\alpha} \rightarrow g g)
= \alpha_s^2
C_g \frac{10}{3}\frac{1}{\pi^3 }
M_{\pi_8^{\alpha}}^3\left(\frac{1}{v}\right)^2
\eeq
where $C_g$ is a dimensionless parameter,
expected to be not too different from one.
It is equal to one for the one family
technicolor model with $N_{TC}=8$. It corresponds to a
phenomenological parameter in our BESS model. The physical $P_8^0$
is a mixed state of $\pi_8^{\alpha}$  and $\pi_8^{\alpha 3}$, and
since
$\pi_8^{\alpha 3}$ does not contribute, we get :
\beq{13}
\Gamma(P_8^0 \rightarrow g g)
=\alpha_s^2 C_g \frac{5}{3}\frac{1}{\pi^3 }
M_{P_8^0}^3 \left(\frac{1}{v}\right)^2
\eeq
For $M_{P_8^0}= 822\ \hbox{GeV}$, the dominant contribution to the
total width $\Gamma_{\hbox{tot}}$ is mainly due to top decay since
$\Gamma(P_8^0 \rightarrow t \bar{t})=60.5 \
{\hbox{GeV}}$ and
$\Gamma(P_8^0 \rightarrow g g)=6.9 \ {\hbox{GeV}}$ (for $C_g=1$
and $m_t=176\ \hbox{GeV}$).

The colored pseudo goldstone bosons being relatively narrow, the
partonic cross section $gg \rightarrow  P_8^0 \rightarrow
t\bar{t}$ reads :
\beq{14}
\frac{\dis d\hat{\sigma}}{\ds d\cos\theta}= \frac{\pi}{4}
\frac{\dis \Gamma(P_8^0 \rightarrow g g)
\Gamma(P_8^0 \rightarrow t\bar{t}) }
{\ds (\hat{s}-M_{P_8^0}^2)^2+M_{P_8^0}^2\Gamma_{P_8^0}^2}
\eeq
The hadronic cross section is obtained after convolution with
the gluon structure function $g(x,M^2)$.

The observable we deal with is the transverse momentum of
the top :
\beq{15}
\ba{ll}
\frac{\dis d\sigma}{\ds dp_T^{\hbox{top}} dy}=
2 p_T^{\hbox{top}}
&\dis \int_{x_{\hbox{min}}}^1dx_1\frac{\dis x_1x_2}
{\ds x_1 -p_T^{\hbox{top}}\sqrt{s} \chi e^y}\\[3mm]
&\left[f_{p_1}^{H_1}(x_1,M^2) f_{p_2}^{H_2}(x_2,M^2)
\frac{\dis
d\hat{\sigma}}{\ds d\hat{t}} (p_1 p_2 \rightarrow t\bar{t})\right]
\ea
\eeq
 where the $f_p^H(x,M^2)$ are the partonic structure functions
evolved at scale $M^2$.
\beq{16}
\ba{ll}
\frac{\dis d\sigma}{\ds dp_T^{\hbox{top}}} &=
2 p_T^{\hbox{top}}\dis \int_{-y_B}^{y_B} dy
 \int_{x_{\hbox{min}}}^1dx_1\frac{\dis x_1x_2}
{\ds x_1 -p_T^{\hbox{top}}\sqrt{s} \chi e^y} g(x_1,4m_t^2)
g(x_2,4m_t^2)\\[3mm]
&\left[\frac{\dis \pi}{\ds 2x_1x_2 s\beta}
\frac{\dis \Gamma(P_8^0 \rightarrow g g)
\Gamma(P_8^0 \rightarrow t\bar{t}) }
{\ds (\hat{s}-M_{P_8^0}^2)^2+M_{P_8^0}^2\Gamma_{P_8^0}^2}
+
\frac{\dis d\hat{\sigma}^{\hbox{SM}}(gg\rightarrow
t\bar{t})}{\ds d\hat{t}}\right]\\[3mm]
\ea
\eeq
where $\hat{t}=-{\hat{s}}\left(1-\beta
\cos\theta\right)/2$, $y$ is the top rapidity,
$$
\matrix{\beta=\sqrt{1-\frac{\dis 4m_t^2}{\dis\hat{s}}} &
x_{\hbox{min}}=
\frac{\dis p_T^{\hbox{top}}\chi e^y}
{\dis \sqrt{s}-p_T^{\hbox{top}}\chi e^{-y}}\cr
&\cr
\chi= \sqrt{1+\frac{\dis m_t^2}{\ds (p_T^{\hbox{top}})^2}} &
x_2=
\frac{\dis x_1p_T^{\hbox{top}}\chi e^{-y}}
{\dis x_1\sqrt{s}-p_T^{\hbox{top}}\chi e^y}\cr}
$$
$\hat{\sigma}^{\hbox{SM}}$ is the QCD cross section \cite{ntl},
slightly modified from the SM expectation due to the
mixing angle $\xi$.

At Tevatron energy the differential cross section  $\frac{\dis
d\sigma}{\ds dp_T^{\hbox{top}}}$ is very small in magnitude since
gluon gluon subprocess is suppressed compared to quark-antiquark
channel. Moreover, the pseudo goldstone $P_8^0$  is too massive to
produce a clear signal distinguishable from the QCD background.

\section{Top pair production from a color-octet massive
vector resonance}

Let us first assume that there is no direct coupling of the massive
octet
vector resonance to quarks. The total width of the massive gauge
boson
$\VV_8^{\alpha}$, assuming five massless quarks and taking into
account only the top quark mass, reads:
\beq{17}
\Gamma(\VV_8^{\alpha})=\frac{1}{12\pi}\frac{\dis g_s'^4}
{\ds g"^2-2g_s'^2} M_{\VV_8}\left[5+\sqrt{1-
\frac{4m_t^2}{M_{\VV_8}^2}}\left(1+\frac{2m_t^2}
{M_{\VV_8}^2}\right)\right]
\eeq
with $g_s'=\frac{\dis g_s g"}{\ds \sqrt{g"^2+2g_s^2}}$.

Top quark production is obtained by coherent sum of the gluon
 and $\VV_8$ s-channel amplitudes. The differential
partonic cross section reads :
$$
\frac{\dis d\hat{\sigma}}{\ds d\hat{t}} =  \frac{\dis \mid
M\mid^2} {\ds 16\pi \hat{s}^2}
$$
with
\beq{18}
\ba{ll}
\mid M\mid^2=& \frac{\dis 4}{\ds 9}g_s^{'4}
\frac{\dis
(m_t^2-\hat{u})^2+(m_t^2-\hat{t})^2+2\hat{s}m_t^2}
{\ds \hat{s}^2}\\[3mm]
&\left[1+\frac{\dis 4g_s^{'4}}{\ds (g"^2-2g_s^{'2})^2}
\frac{\dis \hat{s}^2}
{\ds (\hat{s}-M_{\VV_8}^2)^2+M_{\VV_8}^2
\Gamma_{\VV_8}^2}\right.\\[3mm]
&\left. + \frac{\dis 2g_s^{'2}}{\ds (g"^2-2g_s^{'2})}
\frac{\dis 2\hat{s}(\hat{s}-M_{\VV_8}^2)}
{\ds (\hat{s}-M_{\VV_8}^2)^2+M_{\VV_8}^2
\Gamma_{\VV_8}^2}\right]\\[3mm]
\ea
\eeq
A direct coupling of the colored gauge bosons $\VV_8^{\alpha}$ to
quarks can be considered, extending the construction of
\cite{bess} to the extended BESS model.
 Then the couplings to $q\bar{q}$ pairs become:
\beq{19}
\ba{ll}
g'_s \frac{\dis \lambda^a}{\dis 2} \gamma_{\mu} & \hbox{for the gluon
}\\[3mm]
\frac{\dis \sqrt{2}(b_S g"^2-2g_s^{'2}(1+b_S))}
{\ds 2(1+b_S)\sqrt{g"^2-2g_s^{'2}}} \frac{\dis \lambda^a}{\dis 2}&
\hbox{for the
  }\VV_8^{\alpha}\\[3mm]
\ea
\eeq
In principle we can have two different couplings one for light
quarks, para\-me\-tri\-zed by $b_S$, and one for top and
bottom quarks parametrized by $b_S^\prime$.

The presence of parameters $b_S$ and $b_S^\prime$ different from
zero will modify the expression of the squared matrix element
$\mid M\mid ^2(q\bar{q}\rightarrow g, \VV_8\rightarrow
t\bar{t})$ according to:
\beq{20}
\ba{ll}
\kern -5mm \mid M\mid^2=
& \frac{\dis 4}{\ds 9}
\frac{\dis
(m_t^2-\hat{u})^2+(m_t^2-\hat{t})^2+2\hat{s}m_t^2}
{\ds \hat{s}^2}
\left[ \vpc
g_s^{'4}
+ \frac{\dis \hat{s}^2}{\ds (\hat{s}-M_{\VV_8}^2)^2+M_{\VV_8}^2
\Gamma_{\VV_8}^2}
\right.                                                     \\[5mm]
& \times \frac{\dis \left[b_Sg"^2-2g_s^{'2}(1+b_S)\right]^2
\left[b_S^\prime g"^2-2g_s^{'2}(1+b_S^\prime)\right]^2}
{\ds 4(1+b_S)^2(1+b_S^\prime )^2(g"^2-2g_s^{'2})^2}         \\[5mm]
& + \frac{\dis 2\hat{s}(\hat{s}-M_{\VV_8}^2)}
{\ds (\hat{s}-M_{\VV_8}^2)^2+M_{\VV_8}^2
\Gamma_{\VV_8}^2}                                           \\[5mm]
& \left. \mkern -2mu  \times \frac{\dis g_s^{'2}\left[b_S
g"^2-2g_s^{'2}(1+b_S)\right]
\left[b_S^\prime g"^2-2g_s^{'2}(1+b_S^\prime)\right]}
{\ds 2(1+b_S)(1+b_S^\prime)(g"^2-2g_s^{'2})}
\vpc \right]
\ea
\eeq
The above expression corresponds to an initial state with  light
quarks. For  the process involving bottom  quarks in the initial
state i.e. for the subprocess $b\bar{b}\rightarrow t\bar{t}$ $b_S$
must be replaced by $b_S^\prime $ in the previous relation. We are
now ready to study the increase of the total cross section due to the
colored massive vector resonance and the modification on the
observable
$\frac{\dis d\sigma}{\ds
dp_T^{\hbox{top}}}$.

In this analysis, for simplicity,
we did not include a direct coupling for the axial vector
particles.

\section{Numerical results}

We consider top production at the Tevatron
energy $\sqrt{s}= 1.8\  \hbox{TeV}$. The BESS contribution depends
on two parameters: the mass of the color octet boson and
the strength of its coupling to quarks. We give, in fig.1, the
prediction concerning the transverse momentum spectrum of
the top, for $b_S, b_S^\prime =0$ using two values of the $SU(8)_V$
coupling constants $g"=15$ and $20$, which are consistent with
LEP1 bounds.
The top mass  has been fixed to $170\  \hbox{GeV}$
and the colored vector boson mass to $600\  \hbox{GeV}$. We observe a
slight  excess of events -above the QCD prediction- for
$p_T^{top}\geq 200\  \hbox{GeV}$. This corresponds to the jacobian
peak shifted from  $p_T^{top} \simeq M_{\VV_8}/2$ towards lower
$p_T^{top}$ values due to the heavy top mass. The deviation from
SM expectations is more pronounced for
$g"=15$, leading to an excess of events around
$p_T^{top} \sim 240\  \hbox{GeV}$ by a factor 2-3. There
is no significant increase of the total cross section compared to SM
expectation. When the
colored gauge boson becomes heavier the effect becomes less and less
important. For $M_{\VV_8}\simeq 1\  \hbox{TeV}$ there is no
significant distortion in the transverse momentum spectrum of the
top.

 When direct couplings, i.e. $b_S,b_S^\prime \not= 0$, are allowed,
more drastic
deviations appear. This is illustrated in figures 2 and 3
respectively, for $g"=20$ with $M_{\VV_8} = 600\  \hbox{GeV}$.
In the choice of the parameter $b_S$ and $b_S^\prime$ we restrict
ourselves to values which are allowed by the uncertainty
in our knowledge of the decay $Z$ widths in
hadrons  $\Gamma_h$ and $b\bar b$ $\Gamma_b$.
This is because the new color octet gives vertex corrections
to  $\Gamma_h$ and  $\Gamma_b$ \cite{Zhang}, depending on the choice
of BESS parameters. For instance for $g"=20$,
$M_{{\cal V}_{8}}=400\  \hbox{GeV}$
and $b_S=0.03$ we have $\delta \Gamma_h/\Gamma_h =1.6~10^{-4}$,
 which is one order of magnitude smaller than the experimental value
$\delta \Gamma_h/\Gamma_h$.

Two set of values for
the direct coupling parameters have been chosen in fig.2: $b_S=0$,
$b_S^\prime = -0.03$ and  $b_S=b_S^\prime = -0.03$.
In both cases we observe an excess of events around
$p_T^{top} \sim 240\  \hbox{GeV}$.
 For $b_S=b_S^\prime $ values large and negative, the presence
of the colored vector resonance is clearly manifest
since we would get an increase of the differential cross
section by a factor of $\sim 40$. For
$b_S^\prime = -0.03$ and $b_S=0$,
the existence of a colored massive gauge boson would increase the
differential cross section around $p_T^{top} \sim 240\  \hbox{GeV}$
by roughly a factor of four.  If we choose a smaller $g''$ value
(like $g''=15$) there is a slight increase of the differential cross
section around the jacobian peak. Concerning the total cross
 section its increase depends strongly on the values for
the direct couplings. The greatest discrepancy is observed
when both $b_S$ and $b_S^\prime$ are large and negative,
where we get $\sigma_{tot}=8.24\  \hbox{pb}$. In any case this
value is within $1\sigma$ from the experimental value.
As shown in fig.3 for the sets $b_S=0$,
$b_S^\prime= 0.03$ and  $b_S=b_S^\prime = 0.03$
we get an increase of events
by roughly one order of magnitude around $p_T^{top} \sim 240\
\hbox{GeV}$.

 Deviations become more drastic for lighter $\VV_8$: for
$M_{\VV_8}\simeq 400\  \hbox{GeV}$ the excess of events is displaced
around $p_T^{top}
\sim 100\  \hbox{GeV}$. This is illustrated in fig.4 for $b_S=
b_S^\prime =-0.005$ and  $b_S=b_S^\prime=0.02$ with $g"=20$.  For
$M_{\VV_8} = 600\
\hbox{GeV}$, even for large negative $b_S=b_S^\prime$ values, the
excess of events is within the experimental bounds whereas for
$M_{\VV_8} = 400\
\hbox{GeV}$ the range of direct
couplings still compatible with CDF and D0 experiments
is more restricted for a top mas of $176$ GeV.
The choice of parameters in fig.4
corresponds to the upper and lower limits. In fact, in this case,
for $b_S=b_S^\prime=-0.005$ we get $\sigma_{tot}=8.24\  \hbox{pb}$.
The next to leading QCD corrections \cite{ntl}, not
taken into account in this work, are expected to increase this Born
cross section by roughly 30\% in the same way for the SM
contribution and for the BESS contribution.

 If we increase the top mass up to $200$ GeV,
since the Standard Model value of
the total cross section is smaller, a
larger domain of BESS parameters remains still allowed. As
an example the limits on direct couplings
are  $b_S=b_S^\prime =-0.02$ for  $M_{\VV_8} = 400\
\hbox{GeV}$.

\section{Conclusion}

We have studied top production in the extended
BESS model. At Tevatron energy, since  gluon gluon
subprocess is very small compared to quark antiquark
annihilation subprocess, the contribution of a colored pseudo
goldstone boson is not expected to modify SM
predictions for top production.

The existence of a relatively light massive color-octet
vector resonance $(M_{\VV_8}> 500\ \hbox{GeV})$ leads to an excess of
top production still compatible with present data for a very large
domain of BESS parameters. Therefore the observation of significant
distorsions in the transverse momentum spectrum of the top, and
certainly also of the transverse momentum distribution of the $W$
boson, will be the appropriate way to constrain the BESS model
parameters in the strong  sector and more precisely the direct
couplings of the colored vector resonances to quarks.

 Predictions at LHC will be different since the   gluon gluon
subprocess is dominant. Therefore the anomaly-type contribution is
not expected
to be negligible and top production from decay of pseudo goldstone
bosons has also to be taken into account.

\section*{Figure Captions}

\begin{itemize}
\item[Fig.1] Transverse momentum spectrum of the top at Tevatron
Collider. Full curve: SM prediction. Dashed (resp. dotted)
curve: BESS prediction for $g"=20$ (resp. $g"=15$), $M_{\VV_8} = 600$
GeV and $b_S=b_S^\prime=0$.

\item[Fig.2] Transverse momentum spectrum of the top at Tevatron
Collider. Full curve: SM prediction. Dashed (resp. dotted)
curve: BESS prediction for $b_S=0$ and $b_S^\prime=-0.03$
(resp. $b_S=b_S^\prime=-0.03$), $M_{\VV_8} = 600$ GeV and $g"=20$.

\item[Fig.3] Transverse momentum spectrum of the top at Tevatron
Collider. Full curve: SM prediction. Dashed (resp. dotted)
curve: BESS prediction for $b_S=0$ and $b_S^\prime=0.03$
 (resp $b_S=b_S^\prime=0.03$),
$M_{\VV_8} = 600$ GeV and $g"=20$.

\item[Fig.4] Transverse momentum spectrum of the top at Tevatron
Collider. Full curve: SM prediction. Dashed (resp. dotted)
curve: BESS prediction for  $b_S=b_S^\prime=-0.005$
 (resp. $b_S=b_S^\prime=0.02$), $M_{\VV_8} = 400$ GeV and $g"=20$.

\end{itemize}

\end{document}